# Orbital, spin and valley contributions to Zeeman splitting of excitonic resonances in $MoSe_2$, $WSe_2$ and $WS_2$ monolayers

M. Koperski[1,2,3*], M. R. Molas[1,2], A. Arora[1], K. Nogajewski[1], M. Bartos[1], J. Wyzula[1,4], D. Vaclavkova[1,4], P. Kossacki[2], M. Potemski[1,2]

[1] Laboratoire National des Champs Magnétiques Intenses, CNRS-UJF-UPS-INSA, Grenoble, France
[2] Institute of Experimental Physics, Faculty of Physics, University of Warsaw, Warsaw, Poland
[3] School of Physics and Astronomy, The University of Manchester, Manchester M13 9PL, UK
[4] Department of Experimental Physics, Faculty of Science, Palacký University, Olomouc, Czech Republic
*Maciej.Koperski@manchester.ac.uk

## ABSTRACT

We present a comprehensive optical study of the excitonic Zeeman effects in transition metal dichalcogenide monolayers, which are discussed comparatively for selected materials: $MoSe_2$, $WSe_2$ and $WS_2$. We introduce a simple semi-phenomenological description of the magnetic field evolution of individual electronic states in fundamental sub-bands by considering three additive components: valley, spin and orbital terms. We corroborate the validity of the proposed description by inspecting the Zeeman-like splitting of neutral and charged excitonic resonances in absorption-type spectra. The values of all three terms are estimated based on the experimental data, demonstrating the significance of the valley term for a consistent description of magnetic field evolution of optical resonances, particularly those corresponding to charged states. The established model is further exploited for discussion of magneto-luminescence data. We propose an interpretation of the observed large g-factor values of low energy emission lines, due to so-called bound/localized excitons in tungsten based compounds, based on the brightening mechanisms of dark excitonic states.

**Subject Areas:** Condensed Matter Physics, Nanophysics, Semiconductor Physics

Magneto-optical response of semiconducting transition metal dichalcogenide (sc-TMD) monolayers is commonly investigated to probe the coupling strength of effective out-of-plane angular momentum of carriers in K-valleys with external magnetic field in the limit of 2D confinement. The direct band gap character of sc-TMD monolayers **[1-5]** allows for a comparative study of absorption-type (reflectance and transmission) and photoluminescence processes, both related to the K-valley states. The most obvious manifestation of magnetic coupling is revealed by Zeeman-like splitting of excitonic resonances observed when the magnetic field is applied perpendicularly to the surface of the structure **[6-13]**. The magnitude of the splitting is a quantitative measure of magnetic-field-induced energy shifts of conduction and valence states involved in the optical transitions. As such, the Zeeman-like effect for optical resonances may act as a validity test for theoretical predictions and basic understanding of magnetic field impact on the energy of electronic states in fundamental, K-point, sub-bands **[14-23]**. As for now, the accuracy of developed models is rather poor, so that the explanation of even the simplest observations remains disputable. Especially elusive is the role of so-called valley term, which is responsible for band-type effects sensitive to magnetic field. Here, we aim to deepen the understanding of magneto-optical properties of sc-TMD monolayers by introducing a phenomenological description of linear-with-magnetic-field evolution of conduction and valence states involved in experimentally identified optical transitions.

This simple model includes valley, spin and orbital terms treated as additive components. Three parameters related to these contributions can be estimated by comparatively analyzing polarization-resolved magneto-reflectance spectra of different representatives of sc-TMD family (MoSe$_2$, WSe$_2$ and WS$_2$ are considered in this work). Particularly, the examination of the resonances due to negatively charged excitons provides information on magnetic field splitting of an individual state in the conduction band, which unambiguously demonstrates the significance of the valley term in the interpretation of experimental data and yields an estimation of its value. Further consequences arising from the introduced model are revealed by the analysis of magneto-luminescence spectra, which unveil large g-factor values of lines corresponding to bound/localized excitons combined sometimes with peculiar polarization properties.

The band-edge reflectivity spectra of sc-TMD monolayers are dominated by two robust resonances, known as A and B, which arise due to large spin-orbit splitting **[24]** strongly pronounced in the valence band at the K-point. Both resonances are related to free neutral exciton states and are classified according to the valence sub-band involved in the optical transition **[25-32]**. The lower/higher energy excitons A/B are related to transitions involving a hole from the upper/lower valence band state, respectively. Representative reflectance spectra of MoSe$_2$, WSe$_2$ and WS$_2$ monolayers, measured with circular polarization resolution as a function of the external out-of-plane magnetic field, are presented in **Fig. 1**. In sc-TMD monolayers, the $\sigma^+/\sigma^-$ circular polarization of light is used to probe optical transition in K+/K- valleys, providing a remarkably simple sensitivity to valley degree of freedom **[33-35]**. Upon application of an out-of-plane magnetic field, the splitting between $\sigma^+/\sigma^-$ components (defined as $E_{\sigma^+} - E_{\sigma^-} = g\mu_B B$) appears to be equal for A and B excitons for all three materials within the experimental accuracy and the resulting g-factor values are found to be close to $g \approx -4$. Assuming that Zeeman splitting of an excitonic transition reflects the relative Zeeman shifts of corresponding single-particle subbands, the experimental observation, $g_A = g_B (\approx -4)$, represents one perceptible condition to be fulfilled by Zeeman shifts of individual electronic states. When combined with symmetric evolution of K+/K- states imposed by preservation of time-reversal symmetry, the $g_A = g_B$ condition leads to a 3-term parametrization of linear-with-field contributions to the energy of fundamental conduction and valence subbands. Although different possible sets of these 3 parameters can be formally introduced, we follow here a common wisdom and assume that Zeeman shifts of all electronic subbands are expressed with 3 additive terms, each having a simple physical meaning. Consequently, we introduce the valley ($E_V$), spin ($E_S$) and orbital ($E_{d_2}$) terms which contribute to the energies of spin-orbit-split conduction and valence subbands in the following way:

$$\begin{cases} E_{c,\uparrow}^{K\pm} = \pm E_V + E_S = (\pm g_V + g_S)\mu_B B \\ E_{c,\downarrow}^{K\pm} = \pm E_V - E_S = (\pm g_V - g_S)\mu_B B \\ E_{v,\uparrow}^{K\pm} = \pm E_V + E_S \pm E_{d_2} = (\pm g_V + g_S \pm g_{d_2})\mu_B B \\ E_{v,\downarrow}^{K\pm} = \pm E_V - E_S \pm E_{d_2} = (\pm g_V - g_S \pm g_{d_2})\mu_B B \end{cases}$$

where $E_{c(v),\uparrow(\downarrow)}^{K\pm}$ is the energy of conduction (c) or valence (v), spin-up ($\uparrow$) or spin-down ($\downarrow$) subband in K+ or K- valley, $\mu_B$ is Bohr magneton, B stands for the strength of the magnetic field, and g$_V$, g$_S$, g$_{d2}$ are g-factors corresponding to E$_V$, E$_S$ and E$_{d2}$ terms, respectively. A pictorial representation of how each term independently affects the energy of states in the conduction band (**CB**) and valence band (**VB**) in K+/K- valleys is illustrated in **Fig. 2**. It is important to note, that there exist two possible alignments of spin in the conduction spin-orbit-split subbands.
The energy distance between A and B excitons is mostly defined by a large separation of spin-orbit-split subbands in the valence band whereas a considerable smaller spin-orbit effects in the conduction band cannot be directly inferred from simple absorption/reflectance spectra. Information whether the optically bright A exciton is associated with lower or upper conduction band subband have been however supplied by extensive studies of the photoluminescence response of different sc-TMD monolayers and under various conditions (such as analysis of the spectral response to the in-plane magnetic field and including studies involving unusual configuration of the emitted light with respect to the monolayer plane). It is now rather clear that MoSe$_2$ monolayer is a material with bright ground exciton state **[36,37]**, contrary to WSe$_2$ and WS$_2$ for which the ground exciton state is dark **[37-40]**. The relevant consequence of the existence of two types of materials, tentatively named as 'bright' and 'darkish', is a significant difference in g-factor value for the lower energy conduction band state. The spin and valley terms add up in 'bright' materials and compensate each other in 'darkish' materials, leading to larger magnetic-field-induced splitting of lower conduction state for 'bright' than 'darkish' materials.

In order to investigate the validity of this prediction, we will analyze the magneto-reflectance spectra focusing on charged exciton resonance (CX) (assuming it is a negatively charged state[1]), which appears as a weaker feature below the energy of a free neutral exciton A **[42]**. Typical reflectivity spectra showing CX resonance for MoSe$_2$, WSe$_2$ and WS$_2$ monolayers are presented in **Fig. 3**. We will focus on the interpretation of data for MoSe$_2$ and WS$_2$ materials, considering them as representatives of 'bright' and 'darkish' family. The CX resonance in WSe$_2$ monolayer is very weak and broad in our samples, making it difficult to analyze and interpret the data unambiguously. In the absorption process leading to creation of CX, there is a single electron in the initial state, which at low temperature occupies the lower energy conduction band state. In such case, thermal redistribution of the occupation of this individual electronic state leads to polarization of the CX absorption resonance at higher magnetic fields. The rate of the polarization degree increase is directly indicative of the g-factor value corresponding to this individual state. In Boltzmann approximation to Fermi-Dirac statistics, which is valid if the thermal energy is larger than Fermi energy ($kT \gg E_F$), i.e., when the electron concentration is sufficiently small and effective mass is large, what we expect in our remotely doped monolayers, the polarization degree in external magnetic field is given by a formula:

$$\left|\frac{I_{\sigma^+} - I_{\sigma^-}}{I_{\sigma^+} + I_{\sigma^-}}\right| = tanh\left(\frac{g\mu_B B}{kT}\right) = tanh\left(\frac{(g_S \pm g_V)\mu_B B}{kT}\right)$$

where $I_{\sigma^+}/I_{\sigma^-}$ is the absorption strength of CX resonance detected in $\sigma^+/\sigma^-$ polarization, $g = g_S \pm g_V$ is the g-factor of lower energy conduction band state and might be expressed by a sum or a difference of spin and valley g-factors for bright or dark materials respectively, µ$_B$ is Bohr magneton, B is the value of magnetic field, k is Boltzmann constant and *T* is the effective temperature of the electron gas. In case of MoSe$_2$ monolayer the application of this formula is straightforward, as the CX resonance simply splits with a g-factor $g_{CX} = -4.2 \pm 0.2$ (same value as for neutral exciton resonance $g_X = g_{CX}$) and the polarization degree clearly increases with raising magnetic field. Consequently, one can directly estimate the value $g/T$. We will assume that the temperature of the electron gas is equal to 10 K, taking into account weak heating effects of white light illumination of the sample kept in ~4.2 K exchange gas. Then we obtain a g-factor value for lower energy conduction state of MoSe$_2$ $g = 1.84$. The evolution of the CX resonance in WS$_2$ appears to be much more complicated. The energy dependence on magnetic field for $\sigma^+/\sigma^-$ components is apparently non-linear (see **Fig. 3(f)**). This peculiar evolution originates, as discussed previously in literature **[43-48]**, from contribution of two different CX states to the feature observed in absorption-type spectra. The existence of two CX states in 'darkish' materials is a consequence of spin alignment in conduction sub-bands, which allows an excess electron from either K+ or K- valley to accompany the photo-created electron-hole pair and form a bound CX state. As a result, the final configuration of two conduction electrons may constitute a singlet (*inter*-valley) or triplet (*intra*-valley) state. In 'bright' materials only a singlet state may be expected to be bound due to significant spin-spin (exchange) interaction strength, when two parallel spin states occupy the same valley (hence are characterized by the same pseudo-spin quantum number) for triplet state in bright materials. Therefore, the absence of triplet CX state can be seen as a consequence of Pauli blocking mechanism. The singlet and triplet configurations of conduction electrons in 'bright' and 'darkish' materials are presented in **Fig. 4**. The magnetic field evolution of the CX state in 'darkish' WS$_2$ may be accounted for if two transitions exhibiting the same g-factor value ($g_{CX} = -3.8$; same as for the neutral exciton) value and zero-field splitting of 5 meV are considered **[47]**, in combination with polarization effects imposed by the thermal redistribution of population of the excess electron state. For the singlet state of the charged exciton, the higher energy $\sigma^-$ component accumulates the majority of oscillator strength at higher fields, contrary to the triplet state, for which the lower energy $\sigma^+$ component gets enhanced. Overall, due to small zero field splitting with respect to the linewidths of transitions (~20 meV) the CX features in 'darkish' WS$_2$ gives impression of a single resonance, but upon closer inspection it becomes clear that both singlet and triplet CX states are involved and careful analysis allows estimation of the oscillator strength of $\sigma^+/\sigma^-$ components (see **Fig. 3(g)**) and eventually establish g-factor of the lower energy conduction electron by describing the increase of polarization degree with aforementioned Boltzmann-type formula (see **Fig. 3(h)**). As a result, continuing the assumption of 10 K temperature of electron gas, we obtain the value of the lower conduction state g-factor $g = 1.08$.

---

[1] This assumption is justified by rather large energy distances of charged to neutral exciton observed in our samples. Experiments on gated monolayers **[41]** show binding energies of negatively charged excitons to be considerably larger those of positively charged excitons.

In order to translate the parameters extracted from experiments into the spin and valley terms appearing in our model, we will need to make certain assumptions. One can assume, for instance, that the spin and valley g-factors are equal for MoSe$_2$ and WS$_2$ monolayers, which will lead to a simple linear equation:

$$\begin{cases} g_S + g_V = 1.84 \\ g_S - g_V = 1.08 \end{cases} \rightarrow \begin{cases} g_S = 0.38 \\ g_V = 1.46 \end{cases}$$

An alternative method is to assume that the value of the spin term is the same as for a free electron in vacuum ($g_S = 1$, in our convention) and then obtain the values of the valley terms for 'bright' MoSe$_2$ and 'darkish' WS$_2$ monolayers,

$$\begin{cases} g_V = 2.08, \text{for 'bright' MoSe}_2 \text{ monolayers} \\ g_V = 0.84, \text{for 'darkish' WS}_2 \text{ monolayers} \end{cases}$$

The results of both treatments of experimental data are summarized in **Tab. 1**. Although these estimations are very rough, they quite certainly indicate that the valley term is a substantial part of a cohesive description of the magneto-optical properties of TMDC monolayers and cannot be neglected.

Our description of the magnetic field evolution of the electronic states may be further exploited to shed more light onto the magneto-luminescence spectra. As demonstrated in **Fig. 5**, the PL spectra of MoSe$_2$ monolayers are rather simple, showing neutral and charged exciton lines (X and CX), which split with g-factors values mimicking those observed in absorption-type spectra. Both resonances are polarized in higher fields so that the lower energy ($\sigma^+$) component gains in intensity, most likely due to thermal distributions of excitons between field-split states. The 'darkish' WS$_2$ and WSe$_2$ monolayers exhibit much more complicated PL response. The neutral and charged exciton lines (X and CX) are accompanied by a lower energy multi-peak PL band. It has been argued that the lines forming this band originate, perhaps partially, from recombination of dark exciton states via different brightening mechanisms **[37,49,50]**. A rather outstanding fingerprint of these additional lines is that they exhibit large g-factor values. **Tab. 2** presents the g-factor values for all distinguishable lines in PL spectra of MoSe$_2$, WSe$_2$ and WS$_2$ monolayers. Notably, the lines forming low energy band in WSe$_2$ and WS$_2$ exhibit g-factors from 4 up to 13.5 in absolute value, in most cases significantly larger than g-factors of free excitonic resonances. The origin of enhanced magnetic field splitting of lower energy PL lines in 'darkish' materials remains unclear. However, our description of magnetic effects indicates that it is plausible that larger g-factors are related to recombination of dark exciton states. Conceptually, dark excitons may be brightened due to mechanisms mixing states from K+/K- valleys hence allowing *intra*-valley recombination or one could consider *inter*-valley recombination mediated e. g. by phonons to preserve carrier momentum or defect states which are known to be particularly pronounced in 'darkish' TMDc mono- and multilayers **[51-54]**. Both types of processes are illustrated in **Fig. 6**. Employing our estimation of parameters, we can estimate the g-factor values of such optical transitions in a following way:

$$\begin{cases} g_{inter} = 2(g_{d_2} + 2g_V) \approx 10 \\ g_{intra} = 2(g_{d_2} + 2g_S) \approx 8 \end{cases}$$

assuming the following values of our parameters: $g_{d_2} \approx 2, g_V \approx 1.5, g_S \approx 1$. Notably large g-factors (about -8) have been recently reported for intra valley dark excitons in WSe$_2$ monolayer encapsulated in hBN **[55]**. Our simple consideration takes into account recombination of neutral complexes. More complicated transitions could be realised with charged excitons, involving for instance shake-up process with elevation of an excess electron from lower to upper conduction band state. That could potentially lead to peculiar polarisation properties, when higher energy line gains in intensity at higher fields, as is observed, e. g. for lines A1 in WSe$_2$ monolayer and A2 in WS$_2$ monolayer (see **Fig. 5**).

In summary, we have presented a collection of magneto-optical spectra, based on reflectance and PL measurements, of three sc-TMD compounds: MoSe$_2$, WSe$_2$ and WS$_2$. This data helped us to develop a simple model of three additive terms (valley, spin and orbital contributions), which we used to describe the linear-with-magnetic-field (Zeeman) energy shifts of electronic states in fundamental conduction and valence sub-bands. We have demonstrated the significance of the valley term by analysing the magnetic-field-induced increase of polarisation degree of CX state in absorption-type spectra. Particularly, we have observed a clear signature of singlet and triplet CX states in 'darkish' WS$_2$ monolayers. The values of the g-factors of lower energy conduction band state obtained for MoSe$_2$ and WS$_2$ materials allowed us to estimate the value of the valley term and therefore have a tentative description of the evolution of all the electronic states involved in fundamental optical transitions. As a result of this analysis, we have proposed a reasoning, which sheds light onto the expatiation of large g-factor values for lower energy PL lines in 'darkish' monolayers by attribution them to dark states, which become partially allowed through various brightening mechanisms.

## Acknowledgements

We kindly thank J. Marcus for providing a piece of WSe$_2$ crystal used and acknowledge helpful discussions with A. Slobodeniuk and D. Basko. The authors acknowledge the support from the European Research Council (MOMB project No. 320590), the EC Graphene Flagship project (No. 604391) and the ATOMOPTO project (TEAM programme of the Foundation for Polish Science co-financed by the EU within the ERDFund). The support of LNCMI-CNRS, a member of the European Magnetic Field Laboratory is also acknowledged.

The authors of this manuscript declare no competing financial interest.

# APPENDIX: MATERIALS AND METHODS

## 1. Sample preparation by mechanical exfoliation

Monolayer MoSe$_2$, WSe$_2$ and WS$_2$ flakes were obtained by mechanical exfoliation (polydimethylsiloxane-based technique **[56]**) of bulk crystals in 2H phase. Crystals from different sources were used including commercial suppliers (HQ Graphene). Exfoliated flakes were deposited on Si/SiO$_2$ substrates and initially inspected under an optical microscope to determine the thickness of flakes based on their optical contrast. Eventually, the characteristic PL response from measured flakes unambiguously confirms their monolayer thickness.

## 2. Experimental setup and methods

The magneto-optical experiments were done in high magnetic field facility in Grenoble. Specially designed probes were used with resistive magnets (supplying magnetic field up to 30 T with 50 mm Bohr radius) at low temperatures (via helium exchange gas) to measure reflected or emitted light from the samples. The fiber based set-up allowed focalization of the incoming light to a spot of a about ten micrometers in size. Piezo-positioners were used for *x-y-z* movement of the sample. A halogen lamp provided illumination for reflectance measurements and 514.5 nm Ar$^+$ laser line for photoluminescence investigations. A quarter-waveplate followed by a polarizer was mounted in a fixed position before the detection fiber entrance and $\sigma^+/\sigma^-$ helicity was determined by the magnetic field polarity. The reflectivity contrast was obtained by measuring the spectra on two locations: on the flake and on the nearby Si/SiO$_2$ substrate region, then calculated as:

$$R(E) = \frac{R_{flake}(E) - R_{substrate}(E)}{R_{flake}(E) + R_{substrate}(E)}$$

The reflectivity spectra were reproduced by using a transfer matrix method including Lorentzian contributions to account for resonances observed in the experimental spectra. The data were fitted with such phenomenological curves to establish the energy and oscillator strength of transitions separately in spectra measured in $\sigma^+/\sigma^-$ polarizations.

# FIGURES

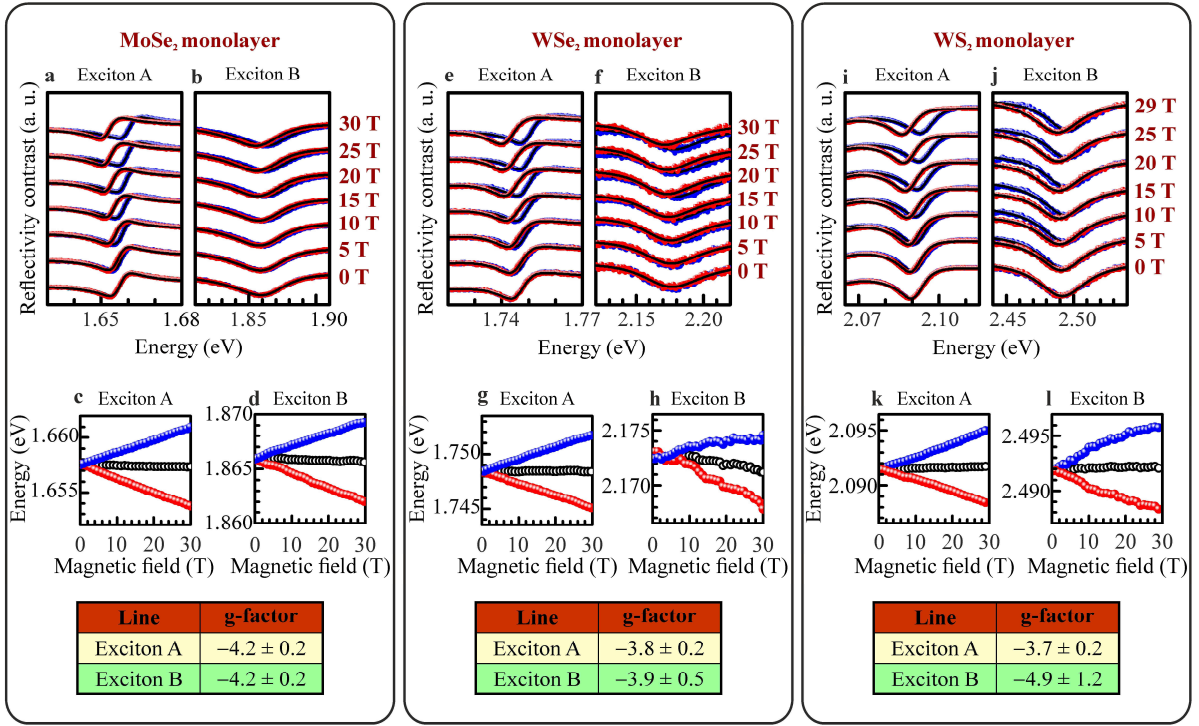

FIG. 1. Magneto-reflectivity spectra, measured with circular polarization resolution (red/blue curves corresponding to $\sigma^+/\sigma^-$ polarization, respectively) under excitation with linearly polarised laser light (514 nm), are shown in the energy range of free neutral exciton resonances A and B for (a-b) MoSe$_2$, (e-f) WSe$_2$ and (i-j) WS$_2$ monolayers. The energies of resonances are obtained by fitting the spectra with curves derived by transfer matrix method including Lorentzian contributions to account for excitonic resonances. The magnetic field evolution of the energy of $\sigma^+$ (red dots) and $\sigma^-$ (blue dots) components are presented for (c-d) MoSe$_2$, (g-h) WSe$_2$ and (k-l) WS$_2$ monolayers. The open circles represent a mean value of the energy of both components $(E_{\sigma^+} + E_{\sigma^-})/2$, to demonstrate that no detectable diamagnetic shift (term $\propto B^2$) is seen. The values of the g-factors for each transition, obtained by fitting the field-dependent energy $E_{\sigma^+} - E_{\sigma^-}$ with linear function $g\mu_B B$, are presented in tables.

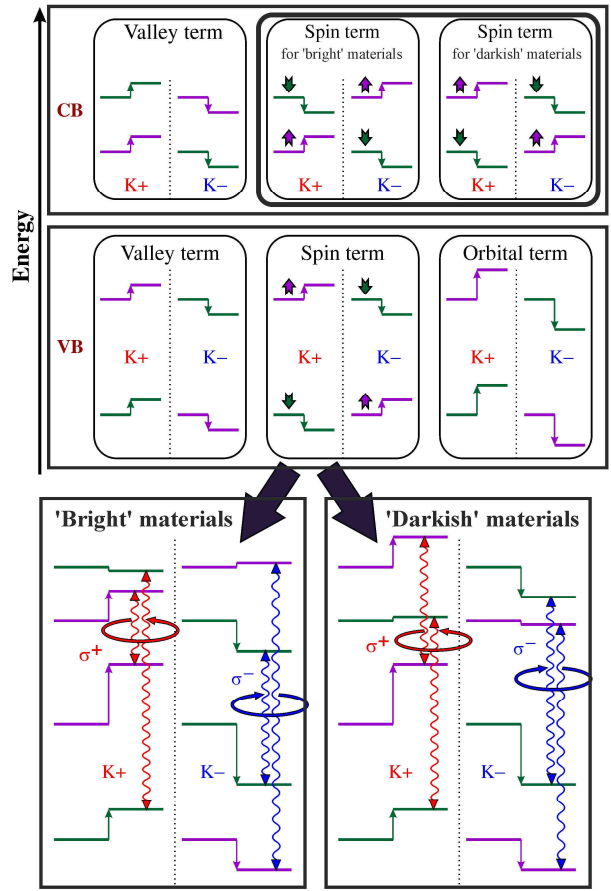

FIG. 2. Scheme of the valley, spin and orbital contributions to the energies of electronic levels is presented for fundamental sub-bands relevant for the optical transitions at K point of the Brillouin zone, which are observed in absorption-type (and emission) spectra. The upper panel shows the evolution of states in the conduction band, the middle panel shows the evolution of states in the valence band and the bottom panel shows the total energy variation (when all contributions are taken into account) for all states involved in optical transition, which are marked with red/blue arrows for $\sigma^+/\sigma^-$ active optical excitations, respectively.

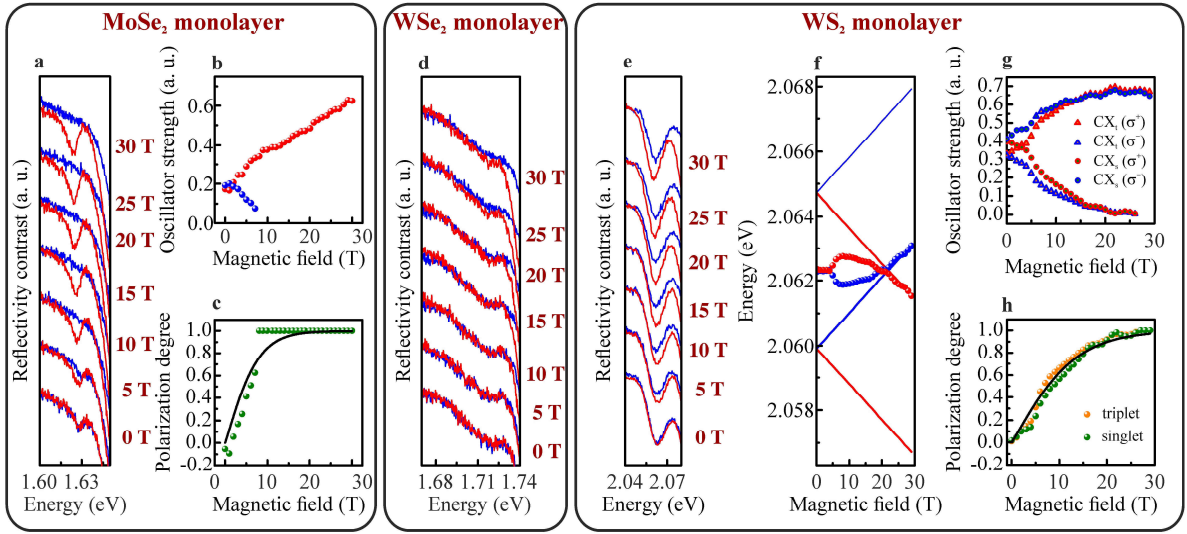

FIG. 3. Magneto-reflectivity spectra of (a) MoSe$_2$, (d) WSe$_2$ and (e) WS$_2$ monolayers measured with circular polarization resolution (red curves for $\sigma^+$ polarization and blue curves for $\sigma^-$ polarization) are presented in the energy region corresponding to CX state (which appears on the lower energy side of the neutral exciton A). The oscillator strength for (b) MoSe$_2$ and (g) WS$_2$ materials is obtained by fitting the curves using transfer matrix method with Lorentzian contributions. In case of MoSe$_2$, a single transition is used which corresponds to a singlet state of CX. For WS$_2$, it was necessary to include two transitions, which we attribute to singlet (CX$_s$) and triplet (CX$_t$) states of CX, with a zero-field splitting (5meV) between them (see main text for further details). For both materials (c and h) the evolution of the polarization degree with magnetic field is presented. The data are fitted with a function (black solid curves) based on Boltzmann occupation of the electronic states, as described in the main text.

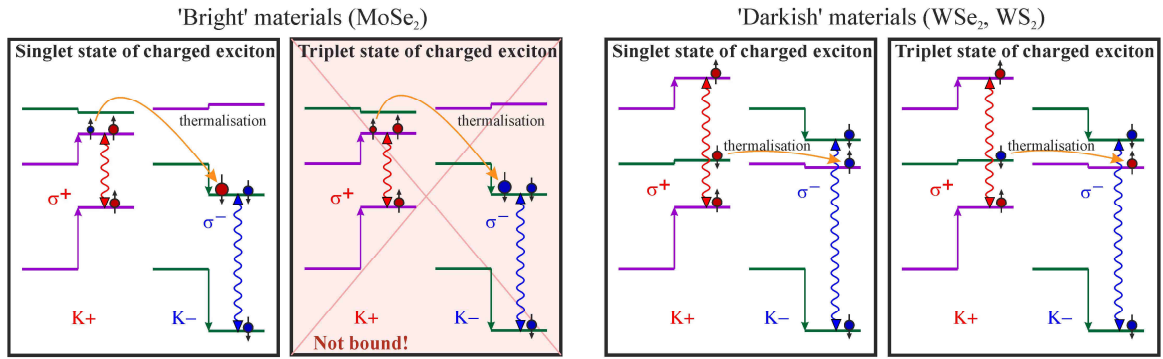

FIG. 4. A scheme presenting possible configurations of electrons and a hole forming CX complexes at low temperatures for 'bright' and 'darkish' sc-TMD materials. The thermalisation effects are demonstrated in the presence of the magnetic field (following the simple additive model described in the main text), leading to an enhancement in higher fields of $\sigma^+/\sigma^-$ red/blue transitions involving carriers marked with red/blue dots. The triplet state for 'bright' materials is most likely not bound due the parallel configurations of electrons occupying the lower energy conduction state in the same valley.

| Nr | Material | Experimental data/assumptions ||||| $g_V$ |
| | | $g_V$ | $g_S$ | $g_{d_2}$ | $g_V - g_S$ | $g_V + g_S$ | |
|---|---|---|---|---|---|---|---|
| 1 | MoSe$_2$ | $g_{V, MoSe_2}$ = $g_{V, WS_2}$ | $g_{S, MoSe_2}$ = $g_{S, WS_2}$ | 2.1 (exp) | | 1.84 (exp) | 1.46 |
| | WS$_2$ | | | 1.9 (exp) | 1.08 (exp) | | 1.46 |
| 2 | MoSe$_2$ | ✗ | 1 (assump.) | 2.1 (exp) | | 1.84 (exp) | 2.08 |
| | WS$_2$ | | | 1.9 (exp) | 1.08 (exp) | | 0.84 |

TAB. 1. Experimentally estimated values of the three terms used to describe the magnetic field impact on the energy of individual states in fundamental sub-bands of sc-TMD monolayers are presented. Two approaches are used, based on different assumptions. In the first case, it is assumed that valley and spin terms are the same for MoSe$_2$ and WS$_2$ materials. In the second case, the spin term is fixed to the value of the free electron in vacuum ($g_S$ = 1). Independently of the approach, the valley term is found to provide a significant contribution to the energy of individual states and its value can be estimated to be $g_V$ = 1.5 ± 0.5.

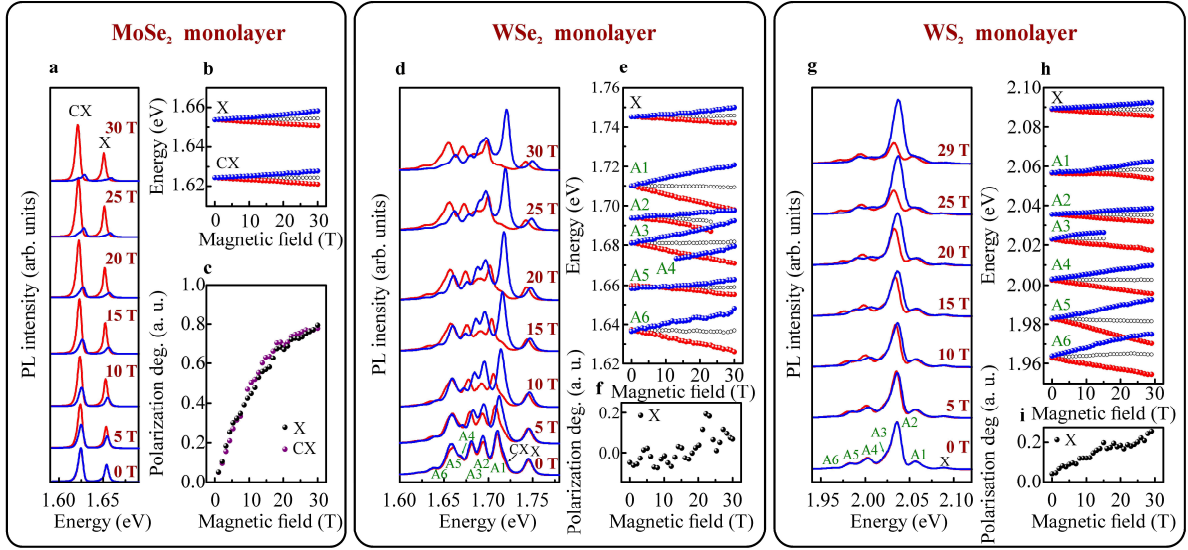

FIG. 5. A collection of magneto-PL spectra detected in circular polarization resolution (red lines and dots for $\sigma^+$ polarization/blue lines and dots for $\sigma^+$ polarization) is presented for (a) MoSe$_2$, (d) WSe$_2$ and (g) WS$_2$ monolayers. The energy of distinguishable lines is presented as a function of magnetic field ((b) for MoSe$_2$, (e) for WSe$_2$ and (h) for WS$_2$ monolayers) and the g-factors are established (see **Tab. 2**). The open circles represented an arithmetic mean value of the energy of $\sigma^+$ and $\sigma^-$ polarized transitions ($E_{mean} = \frac{1}{2}(E_{\sigma^+} + E_{\sigma^-})$). No impact of terms quadratic with magnetic field ($\propto B^2$) is observed. The polarization degree increase with magnetic field for free excitonic states is also presented (for (c) MoSe$_2$, (f) WSe$_2$ and (i) WS$_2$ monolayers), which is indicative of the electron gas effective temperature (about 10 K) under laser light illumination.

| MoSe$_2$ monolayer | |
|---|---|
| Line | g-factor |
| X | −4.2 ± 0.2 |
| CX | −3.7 ± 0.3 |

| WSe$_2$ monolayer | | | |
|---|---|---|---|
| Line | g-factor | Line | g-factor |
| X | −4.3 ± 0.2 | A4 | ∼ −12 |
| A1 | −13.5 ± 0.7 | A5 | −5.3 ± 0.3 |
| A2 | −6.5 ± 0.3 | A6 | −11.4 ± 0.6 |
| A3 | −12.0 ± 0.6 | | |

| WS$_2$ monolayer | | | |
|---|---|---|---|
| Line | g-factor | Line | g-factor |
| X | −4.0 ± 0.2 | A4 | −8.8 ± 0.4 |
| A1 | −4.4 ± 0.2 | A5 | −13.2 ± 0.7 |
| A2 | −4.2 ± 0.2 | A6 | −13.3 ± 0.7 |
| A3 | −8.5 ± 0.4 | | |

TAB. 2. The g-factor values are presented for PL lines of MoSe$_2$, WSe$_2$ and WS$_2$ monolayers. The values for free exciton states correspond well to those obtained from absorption-type experiments. Notably, the g-factors of lower energy PL lines in tungsten based compounds (representatives of 'darkish' sub-class of materials) are significantly larger in absolute value that g-factors of free exciton states (around 4) and reach values as high as 13.

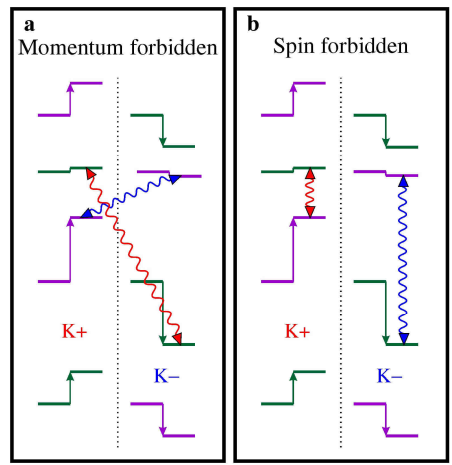

FIG 6. A schematic illustration of forbidden transition in 'darkish' sc-TMD monolayers (such as WSe$_2$ and WS$_2$). (a) The momentum forbidden transitions involve processes with inter-valley recombination, which could be partially allowed considering, e. g. transitions with phonons providing matching momentum. Another type of forbidden transition constitutes (b) inter-valley process, which do not conserve spin. These can be brightened either by mixing of states from different valleys or by considering charged states with recombination realised through spin-flip processes.